\documentclass[10pt,aps,showpacs,nofootinbib,prd,aps,epsf,floats,
               amsmath,amssymb,amsfonts,axodraw,floatfix,graphicx,twocolumn]{revtex4-1}
\usepackage{amsmath, amssymb}
\usepackage{multirow}
\usepackage{paralist}
\newcommand{\mathsym}[1]{{}}

\usepackage{graphicx}% Include figure files
\usepackage{amsmath}
\usepackage{amssymb}
\usepackage{amsmath}
\usepackage{multirow}
\usepackage{paralist}
\usepackage{slashed}
\usepackage{amsfonts}
\usepackage{hyperref} % PDF links
\usepackage{graphicx}% Include figure files
\usepackage{amsmath}
\usepackage{amssymb}
\usepackage{mathrsfs}
\newsavebox{\PSLASH}
 \sbox{\PSLASH}{$p$\hspace{-1.8mm}/}
 
\renewcommand{\theequation}{\thesection.\arabic{equation}}
\newcounter{saveeqn}
\newcommand{\add}{\addtocounter{equation}{1}}
\newcommand{\alphaeqn}{\setcounter{saveeqn}{\value{equation}}%
\setcounter{equation}{0}%
\renewcommand{\theequation}{\mbox{\thesection.\arabic{saveeqn}{\alpha{equation}}}}}
\newcommand{\reseteqn}{\setcounter{equation}{\value{saveeqn}}%
\renewcommand{\theequation}{\thesection.\arabic{equation}}}

 \newsavebox{\notrightarrow}
 \sbox{\notrightarrow}{$\to$\hspace{-4mm}/}
 
 \newsavebox{\PARTIALSLASH}
 \sbox{\PARTIALSLASH}{$\partial$\hspace{-1.6mm}/}
 
 \newsavebox{\ASLASH}
 \sbox{\ASLASH}{$A$\hspace{-2.1mm}/}
 
 \newsavebox{\KSLASH}
 \sbox{\KSLASH}{$k$\hspace{-1.8mm}/}
 
 \newsavebox{\LSLASH}
 \sbox{\LSLASH}{$\ell$\hspace{-1.8mm}/}
 
 \newsavebox{\QSLASH}
 \sbox{\QSLASH}{$q$\hspace{-1.8mm}/}
 
 \newsavebox{\DSLASH}
 \sbox{\DSLASH}{$D$\hspace{-2.2mm}/}
 
 \newsavebox{\DbfSLASH}
 \sbox{\DbfSLASH}{${\mathbf D}$\hspace{-2.8mm}/}
 
 \newsavebox{\DELVECRIGHT}
 \sbox{\DELVECRIGHT}{$\stackrel{\rightarrow}{\partial}$}
 
 \newcommand{\blue}{\IfColor{\textCadetBlue}{}}
\newcommand{\black}{\IfColor{\textBlack}{}}
\newcommand{\red}{\IfColor{\textRed}{}}
\newcommand{\green}{\IfColor{\textOliveGreen}{}}
\newcommand{\lil}{\IfColor{\textRedViolet}{}}

\newcommand{\bs}{\boldsymbol}

\newcommand{\cev}[1]{\reflectbox{\ensuremath{\vec{\reflectbox{\ensuremath{#1}}}}}}
\makeatother
\usepackage[T1]{fontenc}
\usepackage[latin9]{inputenc}
\usepackage{amsmath}
\usepackage{amssymb}
\usepackage{graphicx}
\usepackage{dcolumn}% Align table columns on decimal point
\usepackage{verbatim}

\usepackage{orcidlink}

\begin{document}
\title{Wigner function of a rigidly rotating and magnetized QED plasma}
\author{M. Kiamari~$^{a}$~}\email{mkiamari@ipm.ir}
\author{N. Sadooghi~$^{b}$~\orcidlink{0000-0001-5031-9675}~}\email{Corresponding author: sadooghi@physics.sharif.ir}
\affiliation{$^a$ School of Particles and Accelerators, Institute for Research in Fundamental Sciences (IPM), P.O. Box 19395-5531, Tehran, Iran}
\affiliation{$^b$ Department of Physics, Sharif University of Technology,
P.O. Box 11155-9161, Tehran, Iran}
%%%%%%%%%%%%%%%%%%%%%%%%%%%%%%%%%%%%%%
\begin{abstract}
We determine the Wigner function of a rigidly rotating quantum electrodynamics (QED) plasma in the presence of a constant magnetic field by utilizing the Riemannian normal coordinate approximation, which has been previously proposed in the literature. In this approach, the angular velocity appears only in a specific phase factor, allowing us to compute the point-split fermion two-point correlation function in flat spacetime. To ensure that the fermion correlation function is gauge invariant, we introduce a background gauge field that is fixed to produce a constant magnetic field. Using this Wigner function, we derive the energy-momentum tensor for this medium, which consists of both diagonal and off-diagonal components. By comparing our results with the energy-momentum tensor of an ideal spinful and vortical magnetized fluid, we establish a connection between these components and thermodynamic quantities, such as energy density and different types of pressure. We demonstrate that rigid rotation leads to pressure anisotropy in plasma. Additionally, we compute the associated vector and axial vector currents for this medium, utilizing the previously presented Wigner function. Our results are consistent with existing literature on the subject.
\end{abstract}
\maketitle
%%%%%%%%%%%%%%%%%%%%%%%%%%%%%%%%%%%
\section{Introduction}\label{sec1}
\setcounter{equation}{0}
%%%%%%%%%%%%%%%%%%%%%%%%%%%%%%%%%%%
Current heavy-ion collision (HIC) experiments are focused on producing quark-gluon plasma (QGP) under extreme conditions \cite{rajagopal2018}. Apart from high temperatures reaching up to $10^{12}$ K, extreme conditions include the presence of enormous electromagnetic fields of up to $10^{20}$ Gau\ss~\cite{skokov2009,shen2025} as well as extremely high vorticities reaching up to $10^{22}$ Hz \cite{becattini2017} during the early stages of these collisions. The goal is to replicate the extreme conditions of the early universe within a laboratory setting and to examine the evolution of the QGP created in these collisions.  This process involves studying the transition from an initial out-of-equilibrium state to a later stage of hadronization. In addition to understanding this evolution, it is also crucial to investigate how extreme conditions affect the transport and thermodynamic properties of the QGP.
Chiral magnetic and chiral vortical currents are two types of transport phenomena associated with the presence of external magnetic fields and rotation in the QGP produced after HICs \cite{kharzeev2015}. In recent years, extensive experimental programs have been conducted at the Relativistic Heavy Ion Collider (RHIC) and the Large Hadron Collider (LHC) to investigate currents induced by the chiral magnetic effect (CME) (see \cite{star2025} and references therein). The detection of the chiral magnetic effect (CME) and chiral vortical effect (CVE) in HICs provides strong evidence for understanding the nontrivial topological aspects of Quantum Chromodynamics and the evolution of the magnetic field and angular velocity in these experiments.
\par
The primary theoretical frameworks used to describe the out-of-equilibrium dynamics of the QGP during its initial stages of production are viscous relativistic hydrodynamics and the real-time formalism of thermal field theory. Additionally, relativistic quantum kinetic theory (QKT) serves as an effective approach for exploring the out-of-equilibrium properties of the QGP \cite{hatsuda-book}. The main ingredient of QKT is the Wigner function. In quantum field theory, the Wigner function is defined as the Fourier transform of a point-split fermion two-point correlation function. It replaces the classical partition function, whose dynamics is described by the Boltzmann equation. Being based on a two-point correlation function, the Wigner function captures all quantum corrections that dominate the dynamics of QGP during the initial stages of its formation. In \cite{rischke2017}, the Wigner function for a chirally imbalanced Fermi gas in the presence of background magnetic fields is derived and by making use of this function the CME current is derived. The concept of Wigner function is used to explore the spin polarization of various particles observed in HICs \cite{florkowski2025}. Combining the Wigner function formalism with the fundamental equations of relativistic magnetohydrodynamics, a novel method is presented in \cite{tabatabaee2020} to determine the proper-time evolution of various thermodynamic quantities.
To implement rotation in the Wigner function formalism, it is necessary to formulate it in a curved space.
In \cite{fonarev1993}, the first attempt is made to define the Wigner function for quantum fields coupled to
external gravitational and Yang-Mills fields by using the horizontal lift of covariant derivative in the tangent bundle.  In \cite{huang2018} the chiral kinetic theory in an arbitrary curved spacetime and external magnetic field is formulated by making use of the Wigner function formalism in curved spacetime.
\par
In the present paper, we apply the method described in \cite{fonarev1993} to derive the explicit form of the Wigner function of a rotating QED plasma in the presence of a background magnetic field. Following the assumptions commonly made in theoretical works, we utilize a specific metric that characterizes rigid rotation \cite{mameda2015,fukushima2015}. Based on the findings in \cite{fonarev1993}, we demonstrate that the Wigner function in a rotating frame in the presence of an external magnetic field, consists of two components: A phase factor that incorporates the angular velocity and a kernel that represents the Wigner function in flat spacetime in a background magnetic field.
The latter is described by the point-split two-point function of fermionic fields, which are solutions to the Dirac equation in the presence of a constant magnetic field. To solve this Dirac equation, we employ the Ritus eigenfunction formalism \cite{ritus1972}, leading to a field quantization of magnetized fermions, previously introduced in \cite{taghinavaz2016, tabatabaee2020}.
This method is used in a number of papers \cite{fayazbakhsh2011,  fayazbakhsh2012, fayazbakhsh2013, fayazbakhsh2014, taghinavaz2012}, to study the thermodynamic properties of quark matter in the presence of external magnetic fields. Following same steps as it is demonstrated in \cite{tabatabaee2020}, and taking the phases, including the angular velocity, into account, we derive the Wigner function for a rigidly rotating and magnetized QED plasma. This Wigner function is then employed to calculate the energy-momentum tensor, as well as the vector and axial vector currents in this medium. We demonstrate that the energy-momentum tensor consists of both diagonal and off-diagonal elements. For vanishing angular velocity, the off-diagonal elements disappear, while the diagonal elements are identified with the energy density as well as parallel and perpendicular pressures with respect to the external magnetic field. These results coincide with those presented in \cite{tabatabaee2020}. When the angular velocity is nonvanishing, it is possible to compare the energy-momentum tensor, derived from the above Wigner function with the energy-momentum tensor of an ideal fluid which is spinful, vortical, and magnetized. This model, known as spinful vortical magnetohydrodynamics (SVMHD) was recently introduced in \cite{sedighi2024}. For an observer at rest in the corotating frame, the energy-momentum tensor obtained from SVMHD coincides with that derived from the Wigner function formalism. The energy-momentum tensor presented in this paper, as well as in \cite{sedighi2024}, suggests that, in addition to magnetization \cite{tabatabaee2019}, rigid rotation leads to a pressure anisotropy in the direction perpendicular to the direction of rotation. This anisotropy resembles that found in a viscous fluid and can affect its evolution during expansion \cite{florkowski2011,strickland2018}.
\par
In addition to the energy-momentum tensor, we use the Wigner function of a rotating and magnetized QED plasma to determine the vector and axial vector currents. These currents become zero under two conditions: When the total charge of the plasma is zero, and when the plasma does not exhibit a chiral imbalance. The first condition corresponds to a vanishing chemical potential, $\mu$, while the second refers to a vanishing chiral chemical potential, $\mu_{5}$. The results are consistent with those appearing in the literature \cite{kharzeev2015, abedlou2025}.
\par
The organization of this paper is as follows: In Sec. \ref{sec2}, we begin by briefly reviewing the method presented in \cite{fonarev1993}. Following this, we derive the Wigner function for a rigidly rotating and magnetized QED plasma. In Sec. \ref{sec3}, we utilize this Wigner function to derive the energy-momentum tensor, as well as the vector and axial vector currents of this medium. Section \ref{sec4} is devoted to concluding remarks.  In Appendices \ref{appA} and \ref{appD}, we provide proofs for certain formulas. Appendix \ref{appB} presents the Ritus eigenfunction method, which leads to the solution of the Dirac equation in the presence of a constant magnetic field. Appendix \ref{appC} contains a review of the SVMHD formalism for an ideal fluid. Finally, in Appendix \ref{appE}, some useful integrals are presented.
%%%%%%%%%%%%%%%%%%%%%%%%%%%%%%%%%%%
\section{Wigner function in curved spacetime}\label{sec2}
\setcounter{equation}{0}
\subsection{Review material}\label{sec2A}
%%%%%%%%%%%%%%%%%%%%%%%%%%%%%%
We start with the general definition of the Wigner function in curved spacetime,
\begin{eqnarray}\label{N1}
W_{\rho\sigma}(x,k)=\int d^{4}y\sqrt{-g}~e^{-2ik\cdot y}\langle:\bar{\psi}_{\sigma}(x+y)\psi_{\rho}(x-y):\rangle. \nonumber\\
\end{eqnarray}
Here, $g\equiv \mbox{det}(g^{\mu\nu})$, where $g^{\mu\nu}$ is the metric and $\psi_{\rho},\bar{\psi}_{\sigma}$ are fermionic field operators. In contrast to flat spacetime, where $\psi(x-y)$ is given by $\psi(x-y)=e^{-y\cdot\partial}\psi(x)$, in curved spacetime, we have
\begin{eqnarray}\label{N2}
\hspace{-0.5cm}\psi(x-y)=e^{-y\cdot\mathcal{D}}\psi(x), \quad\bar{\psi}(x+y)=\bar{\psi}(x)e^{y\cdot\mathcal{D}^{\dagger}},
\end{eqnarray}
where the covariant derivative $\mathcal{D}_{\alpha}$ is defined by
\begin{eqnarray}\label{N3}
\mathcal{D}_\alpha \equiv \nabla_{\alpha}- \Gamma _{\alpha\beta }^\gamma y^\beta \partial^y_\gamma.
\end{eqnarray}
Here, $\nabla_{\alpha}\psi\equiv \left(\partial_{\alpha}+\Gamma_{\alpha}\right)\psi$, with the spin connection $\Gamma_{\alpha}\equiv -\frac{i}{4}\omega_{\alpha ij}\sigma^{ij}$, with $\omega_{\alpha ij}\equiv g_{\mu\nu}e^{\mu}_{~i}(\partial_{\alpha}e^{\nu}_{~j}+\Gamma_{\alpha\beta}^{\nu}e^{\beta}_{~j})$ and $\sigma^{ij}=\frac{i}{2}[\gamma^{i},\gamma^{j}]$. The tetrads $e^{\mu}_{~j}$ are defined by $\eta_{ij}=e^{\mu}_{~i}e^{\nu}_{~j}g_{\mu\nu}$ and the Christoffel symbols are given by $\Gamma_{\mu\nu}^{\alpha}\equiv \frac{1}{2}g^{\alpha\sigma}\left(\partial_{\mu}g_{\sigma\nu}+\partial_{\nu}g_{\mu\sigma}-\partial_{\sigma}g_{\mu\nu}\right)$. In these expressions, $g^{\mu\nu}$ and $\eta^{ij}$ are the curved and flat spacetime metrics, respectively. For future purposes, we also define
$\nabla_{\alpha}u^{\beta}=\partial_{\alpha}u^{\beta}+\Gamma_{\alpha\gamma}^{\beta}u^{\gamma}$, with $u^{\alpha}$ a generic four-vector. \\
The second term in the right hand side (rhs) of \eqref{N3} is the horizontal lift.\footnote{A rigorous derivation of the horizontal lift is presented in \cite{huang2018}.} It arises when we bring the Taylor expansion of $\psi(x-y)$,
\begin{eqnarray}\label{N4}
\psi(x-y)=\left( 1 - y^\alpha \nabla_\alpha + \frac{1}{2!} y^\alpha  y^\beta \nabla_\alpha \nabla_\beta \pm\cdots\right)\psi(x),\nonumber\\
\end{eqnarray}
in a compact form $\psi(x-y)=e^{-y\cdot\mathcal{D}}\psi(x)$, with $\mathcal{D}_{\alpha}$ from \eqref{N3}. This is shown by plugging $\mathcal{D}_{\alpha}$ into $e^{-y\cdot\mathcal{D}}\psi(x)$ and using
\begin{eqnarray}\label{N5}
\hspace{-0.5cm}e^{-y\cdot \nabla} = 1 - y^\alpha \nabla_\alpha + \frac{1}{2!} \left(y^\alpha \nabla_\alpha\right)\left(y^\beta \nabla_\beta\right) - \cdots.
\end{eqnarray}
In order to keep the combination of $\bar{\psi}(x+y)\psi(x-y)$ in \eqref{N1} gauge invariant, another modification of $\nabla_{\alpha}$ is necessary. We define,
\begin{eqnarray}\label{N6}
D_\alpha\equiv \mathcal{D}_\alpha-ieA_\alpha,
\end{eqnarray}
with $\mathcal{D}_{\alpha}$ from \eqref{N3} and $A_{a}(x)$ the electromagnetic potential in curved spacetime. By inserting the combination $e^{+y\cdot D^{\dagger}}e^{-y\cdot D}$ into \eqref{N1}, the final form of the Wigner function thus reads
\begin{eqnarray}	\label{N7}
W(x,k)=\int d^{4}y \sqrt{-g}~e^{-2ik\cdot y}\langle:\bar{\psi}(x)
e^{+y\cdot D^{\dagger}}e^{-y\cdot D}
\psi(x):\rangle.\nonumber\\
\end{eqnarray}
By applying the arguments presented in \cite{fonarev1993}, we can reformulate $W(x,p)$ from \eqref{N7} as
\begin{eqnarray}\label{N8}
\lefteqn{\hspace{-0.8cm}
W_{\rho\sigma}(x,k)\big|_{\text{RNC}}= \int d^{4}y \sqrt{-g}~e^{-2ik\cdot y}}\nonumber\\
&&\hspace{-0.5cm}\times\mathcal{Z}^\dagger(x,y)\langle:\bar{\psi}_{\sigma}(x)e^{+y\cdot \partial^{\dagger}}e^{-y\cdot \partial}\psi_{\rho}(x):\rangle\mathcal{Z}(x,-y),
\end{eqnarray}
where $\mathcal{Z}(x,-y)$ is defined by
\begin{eqnarray}\label{N9}
\mathcal{Z}(x,-y)&\equiv&\exp \left( ie\sum_{n=0}^{\infty} \frac{(-1)^{n+1}}{(n+1)!}y^{\alpha_0} \cdots y^{\alpha_n}A_{\alpha_0;\alpha_1 \cdots \alpha_n}\right. \nonumber\\
&&\left.-\frac{e^2}{12}y^{\alpha}y^{\beta}y^{\sigma}[A_{\alpha;\beta},A_{\sigma}]+ \cdots\right).
\end{eqnarray}
In \eqref{N8}, the subscript RNC refers to Riemannian normal coordinate and in \eqref{N9}
$A_{\alpha_{0};\alpha_{i}}\equiv \nabla_{\alpha_{i}}A_{\alpha_{0}}$. It is important to notice that $\psi(x)$ and $\bar{\psi}(x)$ in \eqref{N8} are defined in flat spacetime and the information about curved spacetime is solely included in $\mathcal{Z}(x,y)$. In Appendix \ref{appA}, we outline the proof of \eqref{N8}.
%%%%%%%%%%%%%%%%%%%
\subsection{Wigner function of a rigidly rotating and magnetized Fermi gas}\label{sec2B}
%%%%%%%%%%%%%%%%%%%
As it is described in Sec. \ref{sec1}, the aim of this paper is to determine the Wigner function of a rigidly rotating and magnetized Fermi gas. For simplicity, we assume the gas include positively charged fermions with mass $m$.The rigid rotation around the $z$-axis with the angular velocity $\bs{\Omega}=\Omega \bs{e}_{z}$ is induced by the metric
\begin{equation}\label{N10}
	g_{\mu\nu} =
	\begin{pmatrix}
		1-\rho^{2}\Omega^{2} & \Omega y & -\Omega x & 0\\
		\Omega y & -1 & 0 & 0 \\
		-\Omega x & 0 & -1 & 0 \\
		0 & 0 & 0 & -1
	\end{pmatrix},
\end{equation}
where $\rho^{2}\equiv x^{2}+y^{2}$, and $x$ and $y$ are Cartesian coordinates. For this metric, we have $g=-1$. 
Here, the Greek
indices $\alpha,\beta\in\{0,1,2,3\}$ refer to the general coordinate in
the rotating frame, while the Latin indices  $i,j\in\{t,x,y,z\}$
to the Cartesian coordinate in the local rest frame. Using $g_{\mu\nu}$ from \eqref{N10}, the nonvanishing Christoffel symbols $\Gamma_{\alpha\beta}^{\mu}$ in Cartesian coordinates read
\begin{eqnarray}\label{N11}
\Gamma^{1}_{00} = -\Omega^2 x, &\qquad& \Gamma^{2}_{00} = -\Omega^2 y, \nonumber\\
\Gamma^{2}_{01} = \Gamma^{2}_{10} = \Omega, &\qquad& \Gamma^{1}_{02} = \Gamma^{1}_{20} = -\Omega.
\end{eqnarray}
To introduce a magnetic field aligned in the third direction, $\bs{B}=B\bs{e}_{z}$ with $B>0$, we use the vector potential $A_{a}=(0,-\bs{A})$ in the symmetric gauge, i.e., $A_{a}=\left(0,By/2, -Bx/2,0\right)$. Choosing
\begin{eqnarray}\label{N12}
e^{\mu}_{~a}: e_{~t}^{0} = e_{~x}^{1} = e_{~y}^{2} = e_{~z}^{3} = 1, \;\; e_{~x}^{0} = y \Omega, \;\; e_{~y}^{0} = -x \Omega,\nonumber\\
\end{eqnarray}
the components of $A_{\mu}=e^{a}_{~\mu}A_{a}$ are given by
\begin{eqnarray}\label{N13}
A_{\mu}=(A_0,A_1,A_{2},A_{3})=( -B\Omega \rho^{2}/2, By/2, -Bx/2, 0).\nonumber\\
\end{eqnarray}
Plugging these expressions into $\mathcal{Z}$ from \eqref{N9}, and using the fact that $A_{\alpha}$ is an Abelian gauge field, we obtain
\begin{eqnarray}\label{N14}
\mathcal{Z}(r,-r'/2)&=&\exp \left( \frac{-ieB}{4} \left( y' x - x' y + \Omega t' \rho^{2} \right) \right),\nonumber\\
\mathcal{Z^{\dagger}}(r,r'/2)&=&\exp \left( \frac{ieB}{4} \left( -y' x + x' y - \Omega t' \rho^{2} \right) \right).\nonumber\\
\end{eqnarray}
Combining these expressions, we thus have
\begin{eqnarray}\label{N15}
\mathcal{Z^{\dagger}}(r,r'/2) \mathcal{Z}(r,-r'/2)=e^{\frac{ieB}{2} \left( y' x - x' y + \Omega t' \rho^2\right)}.\nonumber\\
\end{eqnarray}
Here, $r=(t,x,y,z)$ and $r'=(t',x',y',z')$ are defined in flat spacetime. We notice that this specific form for $\mathcal{Z}$ and $\mathcal{Z}^{\dagger}$ is gauge dependent. Plugging \eqref{N15} into \eqref{N8}, we arrive first at
\begin{eqnarray}\label{N16}
W(r,k)= \int d^{4}r' e^{-ik\cdot r'} \langle :\bar{\psi}\left(r+r'/2\right)\psi\left(r-r'/2\right): \rangle e^{i\mathcal{P}}, \nonumber\\
\end{eqnarray}	
with $\mathcal{P}\equiv eB(y'x-x'y+\Omega t'\rho^{2})/2$. Here, $k\cdot r'=k_{a}r^{\prime a}$ with $k_{a}=(k_{t},k_{x},k_{y},k_{z})$. To evaluate $W(r,k)$, we have to use the quantization of $\psi$ and $\bar{\psi}$. According to \cite{tabatabaee2020}, the quantization of fermionic fields $\Psi$ and $\bar{\Psi}$ for the gauge field $\mathcal{A}_{a}=(0,0,-Bx,0)$ is given by
\begin{eqnarray}\label{N17}
\Psi_{\rho}(r)&=&\frac{1}{\sqrt{V}} \sum_{n=0}^{\infty}\sum_{s=1,2} \int \frac{dp_{z}dp_{y}}{(2\pi)^2}\frac{1}{\sqrt{2E_n}}\nonumber\\
& &\times \Biggl\{[\mathbb{P}^{(+)}_{n}(\xi^{(+)})]_{\rho\sigma} u_{s,\sigma}(\tilde{p}_{+}) a^{n,s}_{\bar{p}} e^{-i \bar{p}\cdot \bar{x}} \nonumber\\
&&+ [\mathbb{P}^{(+)}_{n}(\xi^{(-)})]_{\rho\sigma} v_{s,\sigma}(\tilde{p}_{-}) b^{\dagger n,s}_{\bar{p}} e^{+i \bar{p}\cdot \bar{x}} \Biggr\},\nonumber\\
\bar{\Psi}_{\sigma}(r)&=&\frac{1}{\sqrt{V}} \sum_{n=0}^{\infty}\sum_{s=1,2} \int \frac{dp_{z}dp_{y}}{(2\pi)^2}\frac{1}{\sqrt{2E_n}}\nonumber\\
&&\times \Biggl\{a^{\dagger n,s}_{\bar{p}} \bar{u}_{s,\rho}(\tilde{p}_{+})[\mathbb{P}^{(+)}_{n}(\xi^{(+)})]_{\rho\sigma} e^{i \bar{p}\cdot \bar{x}} \nonumber\\
&&+ b^{n,s}_{\bar{p}} \bar{v}_{s,\rho}(\tilde{p}_{-})[\mathbb{P}^{(+)}_{n}(\xi^{(-)})]_{\rho\sigma} e^{-i\bar{p}\cdot \bar{x}} \Biggr\},\quad
\end{eqnarray}
[see Appendix \ref{appB} for more details and necessary notations]. In order to use the above quantization in this paper, where we are working with the symmetric gauge $A_{a}=\left(0,By/2,-Bx/2,0\right)$, we perform an appropriate gauge transformation, $A_{a}=\mathcal{A}_{a}-\partial_{a}\Lambda$, leading to $\Lambda=-Bxy/2$. We thus have $\psi=e^{-ie\Lambda}\Psi=e^{-ieBxy/2}\Psi$. 
\begin{widetext}
Plugging $\psi$ and $\bar{\psi}$ in terms of $\Psi$ and $\bar{\Psi}$ into \eqref{N16}, we obtain
\begin{eqnarray}\label{N18}
W(r,k)= \int d^{4}r' e^{-ik\cdot r'} \langle :\bar{\Psi}\left(r+r'/2\right)\Psi\left(r-r'/2\right): \rangle e^{i\mathcal{W}},
\end{eqnarray}	
with $\mathcal{W}\equiv eB(-2x'y+\Omega t'\rho^{2})/2$. Substituting, at this stage, the quantization relations \eqref{N17} into \eqref{N18}, and using
\begin{eqnarray}\label{N19}
\langle : a_{\bar{\bs{p}}}^{\dagger n,s} a_{\bar{\bs{p}}^{\prime}}^{n',s' }:\rangle &=&+ (2\pi)^{2}V \delta^{2}(\bar{\bs{p}}-\bar{\bs{p}}^{\prime})\delta_{n,n'} \delta_{s,s'} \mathbb{N}_{f}(E_{n}),\nonumber\\
%%%
\langle : b_{\bar{\bs{p}}}^{\dagger n,s} b_{\bar{\bs{p}}^{\prime}}^{n',s' }:\rangle &=& -(2\pi)^{2}V \delta^{2}(\bar{\bs{p}}-\bar{\bs{p}}^{\prime})\delta_{n,n'} \delta_{s,s'} \mathbb{N}_{f}(E_{n}),
\end{eqnarray}
with the Fermi-Dirac distribution function $\mathbb{N}_{f}=(1+e^{\beta E_{n}})^{-1}$ and $E_{n}=\left(m^{2}+2neB+k_{z}^{2}\right)^{1/2}$, as well as  	 
\begin{eqnarray}\label{N20}
\sum_{s} u_{s,\rho}(\tilde{p}_{+}) \bar{u}_{s,\sigma}(\tilde{p}_{+})= \left( \gamma\cdot\tilde{p}_{+}+m \right)_{\rho\sigma},\qquad
\sum_{s} v_{s,\rho}(\tilde{p}_{-}) \bar{v}_{s,\sigma}(\tilde{p}_{-})= \left( \gamma\cdot\tilde{p}_{-}-m\right)_{\rho\sigma},
\end{eqnarray}	
we arrive first at
%%%
\begin{eqnarray}\label{N21}
W(r,k)&=& \sum_{n=0}^{\infty} \int d^{4}r' e^{-ik\cdot r'}e^{i\mathcal{W}}\int \frac{dp_{z}dp_{y}}{(2\pi)^2} \frac{\mathbb{N}_{f}(E_{n})}{2E_{n}} \Biggl\{  [\mathbb{P}^{(+)}_{n}(\zeta^{+})](\gamma\cdot \tilde{p}_{+}+m)[\mathbb{P}^{(+)}_{n}(\zeta'^{+})]  e^{i (E_{n}t' -p_{y}y' -p_{z}z')}\nonumber\\
&& -[\mathbb{P}^{(+)}_{n}(\zeta^{-})](\gamma\cdot \tilde{p}_{-}-m)[\mathbb{P}^{(+)}_{n}(\zeta'^{-})] e^{-i (E_{n}t' -p_{y}y' -p_{z}z')}\Biggr\},
\end{eqnarray}
with
$
\zeta^{\pm}\equiv\left(x\mp p_{y}\ell_{b}^{2}-x'/2\right)/\ell_{b}$ and $\zeta'^{\pm}\equiv\left(x\mp p_{y}\ell_{b}^{2}+x'/2\right)/\ell_{b}$.
Using the definition of $\mathbb{P}_{n}^{(+)}=P_{+}f_{n}^{+}+\Pi_{n}f_{n}^{-}$ from \eqref{appB7} with $P_{\pm}$ from \eqref{appB8} and $f_{n}^{\pm}$ from \eqref{appB9} including the Hermite polynomials and performing the integration over $r'=(t',x',y',z')$ by making use of
\begin{eqnarray}\label{N22}
\int_{-\infty}^{+\infty} dx~e^{-x^{2}}H_{m}(x+y)H_{n}(x+z)=2^{n}\pi^{1/2}m! z^{n-m}L_{m}^{n-m}(-2yz),\quad n\geq m,
\end{eqnarray}
with $L_{m}^{k}(u)$ the associated Laguerre polynomials, and eventually performing the integral over $p_{y}$ and $p_{z}$, we arrive after a lengthy but straightforward computation at the Wigner function of a rigidly rotating and magnetized QED plasma,
%\begin{widetext}
\begin{eqnarray}\label{N23}
\lefteqn{
W(r,k) =2\pi e^{-\frac{\bs{K}_{\perp}^{2}}{eB}}\Biggl\{ \frac{\mathbb{N}_{f}(E_{0})}{E_{0}} \left[ \delta(K_{t}-E_{0})\left(\gamma^{\parallel}\cdot\tilde{p}^{(0)}_{+,\parallel}(-k_{z})+m\right) -\delta(K_{t}+E_{0})\left(\gamma^{\parallel}\cdot\tilde{p}^{(0)}_{-,\parallel}(k_{z})-m\right)\right]P_{+}
}\nonumber\\
	&& +  \sum_{n=1}^{\infty} (-1)^{n} \frac{\mathbb{N}_{f}(E_{n})}{E_{n}} L_{n}\left( \frac{2\bs{K}_{\perp}^{2}}{eB}\right)\left[ \delta(K_{t}-E_{n})\left(\gamma^{\parallel}\cdot\tilde{p}^{(n)}_{+,\parallel}(-k_{z})+m\right) -\delta(K_{t}+E_{n})\left(\gamma^{\parallel}\cdot\tilde{p}^{(n)}_{-,\parallel}(k_{z})-m\right)\right]P_{+}\nonumber\\
	&& -\sum_{n=1}^{\infty} (-1)^{n} \frac{\mathbb{N}_{f}(E_{n})}{E_{n}} L_{n-1}\left( \frac{2\bs{K}_{\perp}^{2}}{eB}\right) \left[ \delta(K_{t}-E_{n})\left(\gamma^{\parallel}\cdot\tilde{p}^{(n)}_{+,\parallel}(-k_{z})+m\right) -\delta(K_{t}+E_{n})\left(\gamma^{\parallel}\cdot\tilde{p}^{(n)}_{-,\parallel}(k_{z})-m\right)\right]P_{-}\nonumber\\
	&& - \sum_{n=1}^{\infty} (-1)^{n} \frac{\mathbb{N}_{f}(E_{n})}{E_{n}} L_{n-1}^{1}\left( \frac{2\bs{K}_{\perp}^{2}}{eB}\right)\sqrt{\frac{2}{n}} \Biggl[ \frac{\left(K_{x}+iK_{y}\right)}{\sqrt{eB}} \left[ \delta(K_{t}-E_{n})\left(\gamma^{\perp}\cdot\tilde{p}^{(n)}_{+,\perp}(-k_{z})\right) -\delta(K_{t}+E_{n})\left(\gamma^{\perp}\cdot\tilde{p}^{(n)}_{-,\perp}(k_{z})\right)\right]P_{-}\nonumber\\
	&&\qquad\qquad+\frac{\left(K_{x}-iK_{y}\right)}{\sqrt{eB}} \left[ \delta(K_{t}-E_{n})\left(\gamma^{\perp}\cdot\tilde{p}^{(n)}_{+,\perp}(-k_{z})\right) -\delta(K_{t}+E_{n})\left(\gamma^{\perp}\cdot\tilde{p}^{(n)}_{-,\perp}(k_{z})\right)\right]P_{+}\Biggr]\Biggr\}.
	\end{eqnarray}
\end{widetext}
Here, $\gamma^{\|}\equiv\left(\gamma^{t},0,0,\gamma^{z}\right)$, $\gamma^{\perp}\equiv\left(0,\gamma^{x},\gamma^{y},0\right)$,
 $\tilde{p}_{\pm,\|}^{(n)}(k_{z})\equiv \left(E_{n},0,0,k_{z}\right)$, and $\tilde{p}_{\pm,\perp}^{(n)}(k_{z})\equiv \left(0,0,\mp\sqrt{2neB},0\right)$. Furthermore, we have $L_{n}^{0}(u)\equiv L_{m}(u)$ and $K_{a}=\left(K_{t},\bs{K}\right)$, with
\begin{eqnarray}\label{N24}
K_{t}=k_{t}-\frac{\Omega \rho^{2}}{2\ell_{b}^{2}},&\qquad&K_{x}=k_{x}+\frac{y}{\ell_{b}^{2}},\nonumber\\
 K_{y}=k_{y}+\frac{x}{\ell_{b}^{2}},&\qquad& K_{z}=k_{z},
\end{eqnarray}
and $\bs{K}_{\perp}^{2}\equiv K_{x}^{2}+K_{y}^{2}$.
 To simplify the combination $P_{\pm}\left(\gamma\cdot\tilde{p}_{\pm}\mp m\right) P_{\pm}$ and $P_{\pm}\left(\gamma\cdot\tilde{p}_{\pm}\mp m\right)P_{\mp}$, we used $P_{\pm}\gamma^{a_{\|}}=\gamma^{a_{\|}}P_{\pm}$, $P_{\pm}\gamma^{a_{\perp}}=\gamma^{a_{\|}}P_{\mp}$, and $P_{\pm}^{2}=P_{\pm}$ as well as $P_{\pm}P_{\mp}=0$. In the next section, we use $W(r,k)$ from \eqref{N23} to determine the energy-momentum tensor and the vector as well as axial vector currents in a rigidly rotating QED plasma in the presence of constant magnetic fields.
%%%%%%%%%%%%%%%%%%%%%%%%%%%%%%%%%%%
\section{Applications}\label{sec3}
\subsection{Energy-momentum tensor of a rigidly rotating magnetized QED plasma}\label{sec3A}
\setcounter{equation}{0}
%%%%%%%%%%%%%%%%%%%%%%%%%
We begin with the Lagrangian density
\begin{eqnarray}\label{E1}
\hspace{-0.5cm}{\cal{L}}=\frac{1}{2}\bar{\psi}\left(i\gamma^{\mu}\vec{D}_{\mu}-m\right)\psi-\frac{1}{2}\bar{\psi}\left(i\gamma^{\mu}\cev{D}_{\mu}-m\right)\psi,\nonumber\\
\end{eqnarray}
with $D_{\mu}$ from \eqref{N6}, including the horizontal lift $\mathcal{D}_{\mu}$ and the gauge field $A_{\mu}$.  The energy-momentum tensor corresponding to $\mathcal{L}$ is given by
\begin{eqnarray}\label{E2}
\mathcal{T}^{\mu\nu}&=&\frac{i}{4}\bar{\psi}\left(\gamma^{\mu}\vec{D}^{\nu}+\gamma^{\nu}
\vec{D}^{\mu}-\gamma^{\mu}\cev{D}^{\nu}-\gamma^{\nu}
\cev{D}^{\mu}\right)\psi\nonumber\\
&&-g^{\mu\nu}{\cal{L}}.
\end{eqnarray}
According to our arguments in Appendix \ref{appD}, the thermal average of $\mathcal{T}^{\mu\nu}$ in terms of the Wigner function reads\footnote{In \cite{huang2020}, another expression for $T^{\mu\nu}$ is presented. In Appendix \ref{appD}, we proof that in RNC used in the present paper, the expression arising in \cite{huang2020} is equivalent with \eqref{E3}.}
\begin{eqnarray}\label{E3}
\lefteqn{T^{\mu\nu}=\langle:\mathcal{T}^{\mu\nu}:\rangle}\nonumber\\
&=&\bigg[\frac{1}{2}\left(\delta_{~\lambda}^{\mu}\delta^{\nu}_{~\sigma}+\delta^{\nu}_{~\lambda}
\delta^{\mu}_{~\sigma}\right)-g^{\mu\nu}g_{\lambda\sigma}\bigg]\int\frac{d^{4}k}{(2\pi)^{4}}k^{\sigma}\mathcal{V}^{\lambda}, \nonumber\\
\end{eqnarray}
with
\begin{eqnarray}\label{E4}
\mathcal{V}^{\lambda}\equiv \mbox{tr}\left(\gamma^{\lambda}W\right).
\end{eqnarray}
Here, we used $g=-1$. Plugging $k^{\sigma}=g^{\sigma\rho}k_{\rho}=g^{\sigma\rho}e^{b}_{~\rho}k_{b}$ and $\mathcal{V}^{\lambda}=e^{\lambda}_{~a}\mathcal{V}^{a}$ into \eqref{E3}, we arrive first at
\begin{eqnarray}\label{E5}
T^{\mu\nu}&=&\left[\frac{1}{2}\left(g^{\nu\rho} e_{a}^{\mu} e^{b}_{\rho} +g^{\mu\rho} e_{a}^{\nu} e^{b}_{\rho} \right) - g^{\mu\nu}\delta_{b}^{a}\right]\int \frac{d^{4}k}{(2\pi)^{4}}~k_{b}\mathcal{V}^{a}.\nonumber\\
\end{eqnarray}
Plugging the Wigner function $W(r,k)$ from \eqref{N23} into $\mathcal{V}^{\lambda}$ and using
\begin{widetext}
\begin{eqnarray}\label{E6}
\lefteqn{\mbox{tr}\left\{\gamma^{a}\left[ \delta(K_{t}-E_{n})\left(\gamma^{\parallel}\cdot\tilde{p}^{(n)}_{+,\parallel}(-k_{z})+m\right) -\delta(K_{t}+E_{n})\left(\gamma^{\parallel}\cdot\tilde{p}^{(n)}_{-,\parallel}(k_{z})-m\right)\right]P_{\pm}\right\}
}\nonumber\\
&&\qquad = 2\left[ \delta(K_{t}-E_{n})\left(\eta^{at}E_{n} +\eta^{az} k_{z}\right) -\delta(K_{t}+E_{n})\left(\eta^{at}E_{n} -\eta^{az} k_{z}\right)\right),\nonumber\\
\lefteqn{\mbox{tr}\left\{
\gamma^{a}	\left[ \delta(K_{t}-E_{n})\left(\gamma^{\perp}\cdot\tilde{p}^{(n)}_{+,\perp}(-k_{z})\right) -\delta(K_{t}+E_{n})\left(\gamma^{\perp}\cdot\tilde{p}^{(n)}_{-,\perp}(k_{z})\right)\right]P_{\mp}
\right\}
}\nonumber\\
&&\qquad =2\left[\delta(K_{t}-E_{n})\left(\eta^{ay} \mp i\eta^{ax} \right) -\delta(K_{t}+E_{n})\left(\eta^{ay} \mp i\eta^{ax} \right)\right] \sqrt{2neB},
\end{eqnarray}
\end{widetext}
and the integrals \eqref{appE1}, we arrive at $T^{\mu\nu}$,
\begin{eqnarray}\label{E7}
\lefteqn{T^{\mu\nu}}\nonumber\\
&=&\left(
	\begin{array}{cccc}
		T^{tt} & y\Omega T^{tt} & -x\Omega T^{tt} & 0 \\
		y\Omega T^{tt} & T^{xx}+ y^{2}\Omega^{2} T^{tt} & -xy\Omega^{2} T^{tt} & 0\\
		-x\Omega T^{tt} & -xy\Omega^{2} T^{tt} & T^{yy}+ x^{2}\Omega^{2} T^{tt} & 0 \\
		0 & 0 & 0 & T^{zz}\\
	\end{array}
	\right),\nonumber\\
\end{eqnarray}
with
\begin{eqnarray}\label{E8}
T^{tt}&=& \frac{eB}{\pi^2} \sum_{n=0}^{\infty} \alpha_n \int_{0}^{\infty} dk_{z}  \frac{\mathbb{N}_{f}(E_{n})}{E_{n}} \left( E_{n}^{2} + m^{2} \right), \nonumber\\
T^{xx}&=& \frac{eB}{\pi^2} \sum_{n=0}^{\infty} \alpha_n \int_{0}^{\infty} dk_{z}  \frac{\mathbb{N}_{f}(E_{n})}{E_{n}} \left( m^{2} + neB \right), \nonumber\\
T^{yy}&=& \frac{eB}{\pi^2} \sum_{n=0}^{\infty} \alpha_n \int_{0}^{\infty} dk_{z}  \frac{\mathbb{N}_{f}(E_{n})}{E_{n}} \left( m^{2} + neB \right), \nonumber\\
T^{zz} &=& \frac{eB}{\pi^2} \sum_{n=0}^{\infty} \alpha_n \int_{0}^{\infty} dk_{z}  \frac{\mathbb{N}_{f}(E_{n})}{E_{n}} \left( k_{z}^{2} + m^{2} \right).
\end{eqnarray}
The energy-momentum tensor $T^{\mu\nu}$ from \eqref{E7} is symmetric and includes, in contrast to the energy-momentum tensor of a nonrotating QED plasma from \cite{tabatabaee2020},
off-diagonal elements depending on $\Omega$. The $T^{at}$ and $T^{ta}$ with $a=1,2,3$ components are interpreted as energy and momentum flux, respectively. Other $T^{ij}, i,j=1,2,3$ are interpreted as shear stress and momentum flux. For $\Omega=0$, we arrive at the energy-momentum tensor in a flat spacetime presented in \cite{tabatabaee2020}.
\par
It is noteworthy that the above result from $T^{\mu\nu}$ is indeed expected from the relation
\begin{eqnarray}\label{E9}
T^{\mu\nu}=e_{~a}^{\mu} e_{~b}^{\nu} T^{ab},
\end{eqnarray}
where $e^{\mu}_{~a}$ are the vierbeins from \eqref{N12} and $T^{ab}$ is the energy-momentum tensor in the local rest frame of the QED plasma presented in \cite{tabatabaee2020}.
\par
In what follows, we compare the above result \eqref{E7} with the energy-momentum tensor arising from ideal SVMHD (spinful vortical MHD) from \cite{sedighi2024} in order to identify different components of \eqref{E7} with thermodynamical quantities. As it is shown in \cite{sedighi2024}, $T^{\mu\nu}$ in a spinful and vortical ideal fluid is given by
\begin{eqnarray}\label{E10}
\lefteqn{T^{\mu\nu}=\epsilon u^{\mu}u^{\nu}-p_{\perp}\Xi^{\mu\nu}+p_{\times} \omega^{\mu}\omega^{\nu}+p_{\|} b^{\mu}b^{\nu}
}\nonumber\\
&&+\frac{B^{2}}{2}\left(u^{\mu}u^{\nu}-\Xi^{\mu\nu}-b^{\mu}b^{\nu}+\omega^{\mu}\omega^{\nu}\right)+2\mu_{b\|}\sigma_{\omega\|}b^{(\mu}\omega^{\nu)},\nonumber\\
\end{eqnarray}
[see Appendix \ref{appC} for necessary notations]. Here, $\epsilon$ is the energy density, $u^{\mu}$ is the fluid velocity, $b^{\mu}$ and $\omega^{\mu}$ are unit vectors in the direction of the magnetic field and rotation, and $p_{\perp}, p_{\times}$, as well as $p_{\|}$ are various types of pressure defined in \eqref{appC8}.Taking the fluid velocity in the local rest frame of the fluid as $u^{a}=(1,\bs{0})$, $u^{\mu}$ in the rotating frame reads
$u^{\mu}=e^{\mu}_{~a}u^{a}$, with $e^{\mu}_{~a}$ from \eqref{N12}. Using this $u^{\mu}$ and the definition of $\omega^{\mu}$ from Appendix \ref{appC}, we get $\omega^{\mu}=(0,0,0,1)$. 
Plugging these expressions into \eqref{E10} and assuming $\mu_{b\|}=\mu_{\omega\|}=0$ as well as $b^{\mu}=(0,0,0,1)$, the energy-momentum tensor is thus given by 
\begin{eqnarray}\label{E11}
T^{\mu\nu}=\left(
	\begin{array}{cccc}
		\epsilon  & y\Omega \epsilon & -x\Omega  \epsilon& 0 \\
		y\Omega\epsilon& P_{\perp}+y^2 \Omega^{2}\epsilon& -x y\Omega ^2\epsilon & 0 \\
		-x\Omega \epsilon& -x y\Omega ^2\epsilon & P_{\perp}+x^2\Omega ^2 \epsilon& 0 \\
		0 & 0 & 0 & P_{\|}\\
	\end{array}
	\right),\nonumber\\
\end{eqnarray}
with $P_{\perp}\equiv p_{0}+B^{2}/2-MB$ and $P_{\|}\equiv p_{0}-B^2/2$.  Here, the magnetization is given by $M=\chi_{m}B$ with $\chi_{m}$ the magnetic susceptibility. The subscripts $\|$ and $\perp$ denote the parallel and perpendicular directions with respect to the direction of the magnetic field and vorticity, $b^{\mu}$ and $\omega^{\mu}$. The choice $\mu_{b\|}=\mu_{\omega\|}=0$ means that there is no net spin in the medium.
As it is described in Appendix \ref{appC}, $B^{2}/2$ terms in $P_{\perp}$ and $P_{\|}$ are the energy densities of the magnetic field.
Comparing \eqref{E7} with \eqref{E11}, it is possible to identify $T^{tt},T^{xx},T^{yy}$, and $T^{zz}$ with thermodynamic quantities, according to
\begin{eqnarray}\label{E12}
T^{tt}=\epsilon, \quad T^{xx}=T^{yy}=P_{\perp},\quad T^{zz}=P_{\|}.\qquad
\end{eqnarray}
It is noteworthy that the definition of $P_{\perp}$ and $P_{\|}$ are expected from \cite{fayazbakhsh2014}. As it is known from \cite{tabatabaee2020}, for $\Omega=0$, $P_{\perp}-P_{\|}=HB$, where $H=B-M$ is the induced magnetic field. Equivalently, in the absence of rotation, we have $HB=\frac{1}{2}(T^{xx}+T^{yy})-T^{zz}$. For $\Omega\neq 0$, an $\Omega$-dependent term is added to $P_{\perp}$ in\eqref{E7}. We thus have,
\begin{eqnarray}\label{E13}
P_{\text{anisotrop}}\equiv\frac{1}{2}(T^{11}+T^{22})-T^{33}=HB+\frac{1}{2}\rho^{2}\Omega^{2}\epsilon,\nonumber\\
\end{eqnarray}
where $\rho^{2}=x^{2}+y^{2}$, is the radius of rigidly rotating cylinder. It thus turns out that apart from magnetization, rigid rotation of the plasma yields a pressure anisotropy, similar to anisotropies arising in a viscous fluid. It is noteworthy that the second term on the rhs of \eqref{E13}, is the pressure corresponding to the centrifugal force of the rotating medium. Additionally, having in mind that the second derivative of the pressure with respect to $\Omega$ yields the moment of inertia of the rotating medium $I$, we arrive at
\begin{eqnarray}\label{E14}
I\equiv \frac{d^{2}P_{\text{anisotrop}}}{d\Omega^{2}}=\rho^{2}\epsilon,
\end{eqnarray}
where $\epsilon$ is the energy density in the corotating frame. 
%%%%%%%%%%%%%%%%%%%%%%%%%%%%%%%%%%%
\subsection{Vector and axial vector currents of a rigidly rotating magnetized QED plasma}\label{sec3B}
We start with the definition of the vector current $j_{V}^{\mu}$ in terms of $\eqref{E4}$,
\begin{eqnarray}\label{E15}
j^{\mu}_{V}= \langle :\bar{\psi}\gamma^{\mu}\psi:\rangle=\int \frac{d^{4}k}{(2\pi)^{4}} \mathcal{V}^{\mu}.
\end{eqnarray}
It is possible to reformulate $\mathcal{V}^{\mu}$ as $\mathcal{V}^{\mu}=e^{\mu}_{~a}\mathcal{V}^{a}$ with vierbeins $e^{\mu}_{~a}$ from \eqref{N12}. Plugging the Wigner function \eqref{N23} into $\mathcal{V}^{\mu}$ and using \eqref{E6}, we arrive first at
\begin{widetext}
\begin{eqnarray*}
j_{V}^{\mu}(r) &= & 4\pi e^{\mu}_{~a}\bigg\{\int \frac{d^{4}k}{(2\pi)^{4}}\frac{\mathbb{N}_{f}(E_{0})}{E_{0}} e^{-\frac{\bs{K}_{\perp}^{2}}{eB}}
\left[\left(\eta^{at}E_{0}+\eta^{az}k_{z}\right)\delta(K_{t}-E_{0}) -\left(\eta^{at}E_{0}-\eta^{az}k_{z}\right)\delta(K_{t}+E_{0})\right]\nonumber \\
&& +\int \frac{d^{4}k}{(2\pi)^{4}}e^{-\frac{\bs{K}_{\perp}^{2}}{eB}} \sum_{n=1}^{\infty} (-1)^{n} \frac{\mathbb{N}_{f}(E_{n})}{E_{n}} \left[ L_{n}\left( \frac{2\bs{K}_{\perp}^{2}}{eB}\right) -  L_{n-1}\left( \frac{2\bs{K}_{\perp}^{2}}{eB}\right) \right]\nonumber\\
&&\qquad \times
  \left[\left(\eta^{at}E_{n}+\eta^{az}k_{z}\right)\delta(K_{t}-E_{n}) -\left(\eta^{at}E_{n}-\eta^{az}k_{z}\right)\delta(K_{t}+E_{n})\right]\nonumber\\
\end{eqnarray*}
\begin{eqnarray}\label{E16}
&& -2\int \frac{d^{4}k}{(2\pi)^{4}} e^{-\frac{\bs{K}_{\perp}^{2}}{eB}} \sum_{n=1}^{\infty} (-1)^{n} \frac{\mathbb{N}_{f}(E_{n})}{E_{n}} L_{n-1}^{1}\left(\frac{2\bs{K}_{\perp}^{2}}{eB}\right)\nonumber\\
&&\qquad \times  \left[\delta(K_{t}-E_{0})+\delta(K_{t}+E_{0})\right]\left[\left(K_{y}+iK_{x}\right)\left(\eta^{ay}-i\eta^{ax}\right)
+\left( K_{y}-iK_{x}\right)\left(\eta^{ay}+i\eta^{ax}\right)\right]\bigg\},
\end{eqnarray}
\end{widetext}
where $K_{a}, a=t,x,y,z$ are defined in \eqref{N24}. The integrand in \eqref{E16} includes terms which are odd in $k_{x},k_{y}$, and $k_{z}$. They vanish after integration over these components. For the integration over $k_{t}$, we use
$\int dK_{t}\left(\delta(K_{t}-E_{0})-\delta(K_{t}+E_{0})\right)=0$ and $\int dK_{t}\left(\delta(K_{t}-E_{n})-\delta(K_{t}+E_{n})\right)=0$. We thus arrive at
\begin{eqnarray}\label{E17}
j_{V}^{\mu}=0.
\end{eqnarray}
Neglecting the effect of the magnetic field, this result is indeed expected from literature \cite{kharzeev2015, abedlou2025}; In \cite{kharzeev2015}, the third component of the vector current $\bs{j}_{V}$ for a dense and chirally imbalanced medium, with chemical potential $\mu$ and axial chemical potential $\mu_{5}$, subjected to an angular velocity $\bs{\Omega}=\Omega\bs{e}_{z}$ is given by
\begin{eqnarray}\label{E18}
j_{V}^{z}=\frac{\mu\mu_{5}\Omega^{2}}{\pi^{2}}.
\end{eqnarray}
In the absence of a chirality imbalance, i.e., for $\mu_{5}$=0, the $z$-component of the vector current vanishes. Other components of $\bs{j}_{V}$ vanish, once the rotation is aligned in the third direction.
\par
Similarly, the axial vector current $j^{\mu}_{A}$ defined by
\begin{eqnarray}\label{E19}
j^{\mu}_{A}= \langle :\bar{\psi}\gamma^{\mu}\gamma^{5}\psi:\rangle=\int \frac{d^{4}k}{(2\pi)^{4}} \mathcal{A}^{\mu},
\end{eqnarray}
with
\begin{eqnarray}\label{E20}
\mathcal{A}^{\mu}\equiv\mbox{tr}\left(\gamma^{\mu}\gamma^{5}W\right).
\end{eqnarray}
Having in mind that $\gamma^{5}$ is invariant under rotation, we have $\mathcal{A}^{\mu}=e^{\mu}_{~a}\mathcal{A}^{a}$. Plugging $W(r,k)$ from \eqref{N23} into \eqref{E20} and using $\mbox{tr}\left(\gamma^{5}\gamma^{a}\gamma^{b}\gamma^{c}\gamma^{d}\right)=4i\epsilon^{abcd}$, we arrive first at
\begin{widetext}
\begin{eqnarray}\label{E21}
\lefteqn{\mbox{tr}\bigg\{\gamma^{5}\gamma^{a}	\left[ \delta(K_{t}-E_{n})\left(\gamma^{\perp}\cdot\tilde{p}^{(n)}_{+,\perp}(-k_{z})\right) -\delta(K_{t}+E_{n})\left(\gamma^{\perp}\cdot\tilde{p}^{(n)}_{-,\perp}(k_{z})\right)\right]P_{\mp}\bigg\}}\nonumber\\
&&\qquad=\frac{\sqrt{2neB}}{2} \left[ \delta(K_{t}-E_{n}) +\delta(K_{t}+E_{n})\right]\left[\mbox{tr}\left( \gamma^{5}\gamma^{a}\gamma^{y}\right) \mp i  \mbox{tr}\left( \gamma^{5}\gamma^{a}\gamma^{y}\gamma^{x}\gamma^{y}\right) \right],\nonumber\\
\lefteqn{\mbox{tr}\bigg\{\gamma^{5}\gamma^{a}\left[ \delta(K_{t}-E_{n})\left(\gamma^{\parallel}\cdot\tilde{p}^{(n)}_{+,\parallel}(-k_{z})+m\right) -\delta(K_{t}+E_{n})\left(\gamma^{\parallel}\cdot\tilde{p}^{(n)}_{-,\parallel}(k_{z})-m\right)\right]P_{\pm}\bigg\}}\nonumber\\
&&\qquad=\pm 2{E_{n} \epsilon^{atxy} \left[ \delta(K_{t}-E_{n}) -\delta(K_{t}+E_{n})\right] +k_{z} \epsilon^{azxy} \left[ \delta(K_{t}-E_{n}) +\delta(K_{t}+E_{n})\right]}.
\end{eqnarray}
\end{widetext}
The expression on the second line vanishes because $\mbox{tr}\left(\gamma^{5}\gamma^{a}\gamma^{y}\right)=0$ and $\mbox{tr}\left(\gamma^{5}\gamma^{a}\gamma^{y}\gamma^{x}\gamma^{y}\right)=0$. Inserting the expression on the fourth line into \eqref{E19} and integrating over $k^{a}$, we obtain
\begin{eqnarray}\label{E22}
j_{A}^{\mu}=0.
\end{eqnarray}
This is because in the first term, $\int dK_{t}[\delta(K_{t}-E_{n})-\delta(K_{t}+E_{n})]=0$ and in the second term, the integration over an odd function in $k_{z}$ vanishes. The above result is expected from literature:  In \cite{kharzeev2015}, the axial vector current of a rotating plasma with nonvanishing $\mu$ and $\mu_{5}$ is given by
\begin{eqnarray}\label{E23}
j_{A}^{z}=\left(\frac{T^{2}}{6}+\frac{\mu^{2}+\mu_{5}^{2}}{2\pi^{2}}\right)\Omega,
\end{eqnarray}
and $j_{A}^{x}=j_{A}^{y}=0$. Here, the angular velocity is assumed to be directed in the third direction. The term $T^{2}$ has an ambiguous origin. Whereas in \cite{kharzeev2015}, it is indicated that this term arises from gravitational anomaly, it was possible to determine it in \cite{abedlou2025} in the framework of thermal field theory. Neglecting the effect of magnetic field and dropping the term proportional to $T^{2}$, and setting $\mu=0$ as well as $\mu_{5}=0$, we obtain $j_{A}^{z}=0$ which is compatible with our result \eqref{E22} arising from the Wigner function \eqref{N23}, according to \eqref{E19}.
%%%%%%%%%%%%%%%%%%%%%%%%%%%%%%%%%%%
\section{Concluding remarks}\label{sec4}
%%%%%%%%%%%%%%%%%%%%
Following the method introduced in \cite{fonarev1993}, we calculated the Wigner function $W(r,k)$ for a rigidly rotating QED plasma in the presence of a constant magnetic field. By employing the horizontal lift and applying the RNC approximation, we demonstrated that the angular velocity $\Omega$ appears only in a phase factor,  and the kernel of the Wigner function, which is formulated in the corotating frame, is described by a point-split two-point function of magnetized fermions and is independent of $\Omega$. We utilized field quantization for magnetized fermions to derive $W(r,k)$ for nonzero $\Omega$. This quantization was previously applied to determine the Wigner function when $\Omega=0$ and in other contexts \cite{taghinavaz2016}.
Using the Wigner function for a rotating and magnetized plasma, we determined in Sec. \ref{sec3A}, the energy-momentum tensor $T^{\mu\nu}$ for this medium. As expected, when $\Omega=0$, the off-diagonal elements of $T^{\mu\nu}$, which represent shear stress and momentum flux, vanish. The diagonal elements correspond the energy density as well as parallel and perpendicular pressures, denoted by $P_{\perp}$ and $P_{\|}$. In this case, we have $T^{xx}=T^{yy}=P_{\perp}=P_{\|}+HB$, where $H=B-M$ is the induced magnetic field and $M$ is the magnetization of the fluid.
However, for nonvanishing $\Omega$, we found that $T^{11}\neq T^{22}$. According to \eqref{E13}, in addition to $HB$, another term proportional to $\rho^{2}\Omega^{2}\epsilon$ emerges, which describes the pressure anisotropy induced by rigid rotation.\footnote{Here, $\rho$ is the radius of the rigidly rotating cylinder. } Using this result, we showed that the moment of inertia of the rotating medium is given by $\rho^{2}\epsilon$, where $\epsilon$ is the energy density in the corotating frame and is given by $T^{tt}$ from \eqref{E8}.
\par
In addition to the energy-momentum tensor, we also determined the vector and axial vector currents, $j_{V}^{\mu}$ and $j_{A}^{\mu}$, by utilizing the Wigner function $W(r,k)$. As expected, in a medium with zero chemical potential $\mu$ and axial chemical potential $\mu_{5}$, both $j_{V}^{\mu}$ and $j_{A}^{\mu}$ vanish. It would be intriguing to employ the method presented in this paper, to determine the Wigner function of a medium with chiral imbalance. The results arising for $T^{\mu\nu}$, $j_{V}^{\mu}$, and $j_{A}^{\mu}$ will then demonstrate the interplay between the external magnetic field and the rotation in these quantities. We postpone this work to our future publications.
%%%%%%%%%%%%%%%%%%%%%%%
\section{Acknowledgments}
The authors thank A. Shojai Baghini for the valuable discussions that contributed to the formulation of \eqref{E9}.
\begin{appendix}
\section{Proof of \eqref{N8}}\label{appA}
\setcounter{equation}{0}
%%%%%%%%%%%%%%
To prove \eqref{N8}, let us consider $e^{-y\cdot D}\psi(x)$ in \eqref{N7}. Inserting $e^{y\cdot\partial}e^{-y\cdot \partial}=1$ and $e^{y\cdot\mathcal{D}}e^{-y\cdot \mathcal{D}}=1$, arising from $[y\cdot\mathcal{D},y\cdot\mathcal{D}]=0$, into this combination, we first arrive at
\begin{eqnarray}\label{appA1}
e^{-y\cdot D}\psi(x) = \mathcal{Z}(x,-y)\mathcal{P}(x,-y)\psi(x-y),
\end{eqnarray}
with $\psi(x-y)=e^{-y\cdot\partial}\psi(x)$, $\mathcal{Z}(x,-y)\equiv e^{-y\cdot D}e^{y\cdot\mathcal{D}}$, and  $\mathcal{P}(x,-y)\equiv e^{-y\cdot \mathcal{D}}e^{y\cdot \partial}$. Then, using
\begin{eqnarray}\label{appA2}
-y\cdot (D-\mathcal{D}) &=& -iey\cdot A, \nonumber\\
\frac{1}{2}[-y\cdot D,y\cdot \mathcal{D}] & =& \frac{ie}{2} y^\mu y^\nu \nabla_\nu A_\mu.
\nonumber\\
\frac{1}{12}[-y\cdot D,[-y\cdot D,y\cdot \mathcal{D}]] &=&-\frac{ie}{12}y^\alpha y^\mu y^\nu \nabla_\alpha \nabla_\nu A_\mu\nonumber\\
&&+\frac{e^2}{12}y^\alpha y^\mu y^\nu[A_\alpha, \nabla_\nu A_\mu], \nonumber\\
-\frac{1}{12}[y\cdot \mathcal{D},[-y\cdot D,y\cdot \mathcal{D}]]&=&-\frac{ie}{12}y^\alpha y^\mu y^\nu \nabla_\alpha \nabla_\nu A_\mu, \nonumber\\
\end{eqnarray}
we obtain $\mathcal{Z}(x,-y)$ from \eqref{N9} and
\begin{eqnarray}\label{appA3}
\mathcal{P}(x,-y)&=&\exp \left( -y^\mu\Gamma_\mu + y^\mu y^\nu \Gamma^\lambda _{\mu\nu }\left(\frac{1}{2} \partial_\lambda +\partial^y_\lambda\right)\right.\nonumber\\
&&\left. - y^\mu y^\nu y^\gamma \nabla_\gamma \Gamma^\lambda_{\mu\nu} \left(\frac{1}{3} \partial_\lambda + \frac{1}{2}\partial^y_\lambda \right) + \cdots \right).\nonumber\\
\end{eqnarray}
As it turns out, in the RNC approximation, $\mathcal{P}(x,-y)=1$, and the final form of the Wigner function in this system is given by \eqref{N8}.
%%%%%%%%%%%%%%%%%%%%%%%%
\section{Fermions in an external magnetic field}\label{appB}
\setcounter{equation}{0}
%%%%%%%%%%%%%%
In Sec. \ref{sec2}, the Wigner function of a rigidly rotating and magnetized Fermi gas is computed.
To arrive at \eqref{N23}, we used the solution of the Dirac equation in the presence of external magnetic field in a nonrotating frame. In this appendix, we present the solution of the Dirac equation in the presence of external magnetic fields using the Ritus eigenfunction method \cite{ritus1972}.
\par
We begin with the Dirac equation for a charged fermion with mass $m_{q}$ and charge $qe$,
\begin{eqnarray}\label{appB1}
\left(\gamma\cdot\Pi^{(q)}-m_{q}\right)\Psi^{(q)}=0,
\end{eqnarray}
with $\Pi_{a}^{(q)}\equiv i\partial_{a}+eq \mathcal{A}_{a}$. The choice of the gauge field $\mathcal{A}_{a}=(0,0,-Bx,0)$ leads to a magnetic field aligned in the third spatial direction, $\bs{B}=B\bs{e}_{z}$. To solve \eqref{appB1}, we use the ansatz $\Psi_{\kappa}^{(q)}=\mathbb{E}_{n,\kappa}^{(q)}u(\tilde{p}^{(\kappa)}_{q})$ for the positive ($\kappa=+1$) and negative ($\kappa=-1$) energy solutions. Here, $\mathbb{E}_{n,\kappa}$ arises from the Ritus eigenfunction relation,
\begin{eqnarray}\label{appB2}
\left(\gamma\cdot \Pi^{(q)}\right)\mathbb{E}_{n,\kappa}^{(q)}=\kappa\mathbb{E}_{n,\kappa}^{(q)}\left(\gamma\cdot \tilde{p}_{q}^{(\kappa)}\right).
\end{eqnarray}
In $\mathbb{E}_{n,\kappa}$, $n$ labels the Landau levels in the external magnetic field.
Plugging the above ansatz into \eqref{appB1} and utilizing \eqref{appB2}, it turns out that $u(\tilde{p}^{(\kappa)_{q}})$ satisfies the free Dirac equation
\begin{eqnarray}\label{appB3}
&&\left(\gamma\cdot \tilde{p}_{q}^{(+)}-m_{q}\right)u(\tilde{p}_{q}^{(+)})=0,\nonumber\\
&&\left(\gamma\cdot \tilde{p}_{q}^{(-)}+m_{q}\right)v(\tilde{p}_{q}^{(-)})=0.
\end{eqnarray}
As it is demonstrated in \cite{fayazbakhsh2011,  fayazbakhsh2012, fayazbakhsh2013, fayazbakhsh2014, taghinavaz2012, taghinavaz2016}, the Ritus momentum for a particle with charge $qe$ reads
\begin{eqnarray}\label{appB4}
\tilde{p}^{\mu}_{q,\kappa}=(E_{n}^{(q)}, 0,-\kappa s_{q}\sqrt{2n|qeB|},p_{z}),
\end{eqnarray}
with $s_{q}\equiv \mbox{sgn}(qeB)$, and the Ritus eigenfunction $\mathbb{E}_{n,\kappa}^{(q)}$ is given by
\begin{eqnarray}\label{appB5}
\mathbb{E}_{n,\kappa}^{(q)}(\xi_{\kappa}^{s_q})=e^{-i\kappa\bar{p}\cdot\bar{x}}\mathbb{P}_{n,\kappa}^{(q)}(\xi_{\kappa}^{s_q}),
\end{eqnarray}
with
$\bar{p}_{a}\equiv \left(p_{t},0,p_y,p_z\right)$, $\bar{x}^{a}\equiv \left(t,0,y,z\right)$, and
$\xi^{s_{q}}_{\kappa}\equiv (x-\kappa s_{q}p_{y}\ell_{B}^{2})/\ell_{b}$, where $\ell_{b}\equiv |qeB|^{-1/2}$. In addition, the on the mass-shell relation $\tilde{p}^{2}=m_{q}^{2}$ yields the energy $E_{n}^{(q)}$ of charged fermions in the presence of a constant magnetic field $\bs{B}$,
\begin{eqnarray}\label{appB6}
E_n^{(q)}=\left(2n|qeB|+p_z^2+m_{q}^2\right)^{1/2}.
\end{eqnarray}
In \eqref{appB5}, $\mathbb{P}_{n,\kappa}^{(q)}$ is defined by
\begin{eqnarray}\label{appB7}
\mathbb{P}_{n}^{(q)}(\xi_{\kappa}^{s_q})\equiv P_{+}^{(q)}f_{n}^{+s_q}(\xi^{s_q}_{\kappa})+\Pi_{n}P_{-}^{(q)}f_{n}^{-s_q}(\xi_{\kappa}^{s_q}),
\end{eqnarray}
where the factor $\Pi_{n}\equiv 1-\delta_{n,0}$ controls the spin degeneracy in the LLL, and the projectors $P_{\pm}^{(q)}$ reads
\begin{eqnarray}\label{appB8}
P_{\pm}^{(q)}\equiv \frac{1\pm is_q \gamma_1\gamma_2}{2},
\end{eqnarray}
with $P_{\pm}^{(+)}=P_{\pm}$, $P_{\pm}^{(-)}=P_{\mp}$ and $P_{\pm}\equiv\left(1\pm i\gamma_1\gamma_2\right)/2$. The functions $f_{n}^{\pm s_q}(\xi_{\kappa}^{s_q})$, appearing in \eqref{appB7} are given by
\begin{eqnarray}\label{appB9}
\begin{array}{rclcrcl}
f_n^{+s_q}&\equiv&\Phi_{n}(\xi_{\kappa}^{s_q}),&&n&=&0,1,2,\cdots,\\
f_{n}^{-s_q}&\equiv& \Phi_{n-1}(\xi_{\kappa}^{s_q}),&&n&=&1,2,\cdots,
\end{array}
\end{eqnarray}
with $\Phi_{n}$ including Hermite polynomials $H_{n}(z)$
\begin{eqnarray}\label{appB10}
\Phi_{n}(\xi_{\kappa}^{s_q})\equiv a_n\exp\left(-\frac{(\xi_{\kappa}^{s_q})^{2}}{2}\right)H_n(\xi_{\kappa}^{s_q}).
\end{eqnarray}
Here, $a_{n}$ is the normalization factor given by $a_n\equiv \left(2^{n}n!\sqrt{\pi}\ell_{b}\right)^{-1/2}$. According to \cite{taghinavaz2016}, the quantization relation of a positively charged fermion with mass $m\equiv m_{+}$ is given by \eqref{N17}, with $\bar{\bs{p}}=(0,p_y,p_z)$, $E_n\equiv E_n^{(+)}$ defined in \eqref{N5}, $\tilde{p}_{\kappa}\equiv \tilde{p}_{+,\kappa}$ and  $\xi^{\kappa}\equiv \xi^{+}_{\kappa}$ for $\kappa=\pm 1$. Here, $a_{\bs{\bar{p}}}^{n,s}, a_{\bs{\bar{p}}}^{\dagger n,s}$ and $b_{\bs{\bar{p}}}^{n,s},b_{\bs{\bar{p}}}^{\dagger n,s}$ are two sets of creation and annihilation operators satisfying the commutation relations
\begin{eqnarray}\label{appB12}
\hspace{-0.3cm}\{a_{\bs{\bar{p}}}^{n,s}, a_{\bs{\bar{q}}}^{\dagger m,s^{\prime}}\}=(2\pi)^{2}V\delta^{2}\left(\bar{\bs{p}}-\bar{\bs{q}}\right)
\delta_{s,s^{\prime}}\delta_{n,m},\nonumber\\
\hspace{-0.3cm}\{b_{\bs{\bar{p}}}^{n,s}, b_{\bs{\bar{q}}}^{\dagger m,s^{\prime}}\}=(2\pi)^{2}V\delta^{2}\left(\bar{\bs{p}}-\bar{\bs{q}}\right)
\delta_{s,s^{\prime}}\delta_{n,m},
\end{eqnarray}
and the spinors $u_{s,\alpha}, \bar{u}_{s,\alpha}$ as well as $v_{s,\alpha}, \bar{v}_{s,\alpha}$ satisfying the spin summation formula \eqref{N20}.
%%%%%%%%%%%%%%%%%%%%%%%%
\section{The formulation of SVMHD}\label{appC}
\setcounter{equation}{0}
%%%%%%%%%%%%%%
 In this appendix, we outline the formulation of an ideal relativistic MHD of a spinful and vortical fluid, that is originally introduced in \cite{sedighi2024}. In order to fix our notations, we focus, in particular, on the formulation of the energy-momentum tensor corresponding to ideal SVMHD.
\par
A relativistic charged fluid described by SVMHD is governed by the following conservation laws,
\begin{eqnarray}\label{appC1}
\partial_{\mu}J^{\mu}=0,\quad\partial_{\mu}T^{\mu\nu}=0,\quad \partial_{\mu}J^{\mu\nu\rho}=0.
\end{eqnarray}
Additionally, the homogeneous Maxwell equation $\partial_{\mu}\tilde{F}^{\mu\nu}=0$ is to be taken into account. Here, $\tilde{F}_{\mu\nu}=\frac{1}{2}\epsilon_{\mu\nu\rho\sigma}F^{\rho\sigma}$ is the dual tensor to the (electromagnetic) field strength tensor $F^{\rho\sigma}$. Moreover,  $J^{\mu}, T^{\mu\nu}$, and $J^{\mu\nu\rho}=L^{\mu\nu\rho}+\Sigma^{\mu\nu\rho}$ are the electromagnetic current, energy-momentum, and total angular momentum tensors. The total angular momentum tensor consists of an orbital $L^{\mu\nu\rho}$ and a spin tensor $\Sigma^{\mu\nu\rho}$. The orbital angular momentum $L^{\mu\nu\rho}$ is given by $L^{\mu\nu\rho}=x^{\nu}T^{\mu\rho}-x^{\rho}T^{\mu\nu}$ with $T^{\nu\nu}$ consisting of a symmetric $T^{(\mu\nu)}$ and an antisymmetric $T^{[\mu\nu]}$ part. The latter satisfies
\begin{eqnarray}\label{appC2}
\partial_{\mu}\Sigma^{\mu\nu\rho}=-2T^{[\nu\rho]},
\end{eqnarray}
that arises from \eqref{appC1}. The thermodynamical quantities describing this fluid are defined by
\begin{eqnarray}\label{appC3}
u_{\mu}J^{\mu}&=&n,\nonumber\\
u_{\nu}T^{(\mu\nu)}&=&\left(\epsilon+B^2/2\right)u^{\mu},\nonumber\\
u_{\mu}\Sigma^{\mu}_{~\nu\rho}&=&-\sigma_{\nu\rho}.
\end{eqnarray}
Here, $u^{\mu}$ is the fluid velocity, given by $u^{\mu}=\gamma(1,\bs{v})$ with
$\gamma=(1-\bs{v}^{2})^{-1/2}$. Moreover, $n$ is the number density, $\epsilon$ is the
energy density of the fluid, $B^{2}/2$, with $B$ the magnetic field strength, the energy density of the magnetic field,  and $\sigma^{\mu\nu}$ the spin density tensor. Using the first law of thermodynamics,
\begin{eqnarray}\label{appC4}
Tds=d\epsilon-\frac{1}{2}\mu^{\mu\nu}d\sigma_{\mu\nu}-\mu dn+MdB,
\end{eqnarray}
we obtain appropriate definitions for the inverse temperature $\beta=T^{-1}$, the spin potential
$\mu^{\mu\nu}$, the chemical potential $\mu$, and the magnetization $M$
\begin{eqnarray}\label{appC5}
\beta&=&\left(\frac{\partial s}{\partial\epsilon}\right)_{\sigma^{\mu\nu},n, B}, \qquad
\beta\mu^{\mu\nu}=-2\left(\frac{\partial s}{\partial\sigma_{\mu\nu}}\right)_{\epsilon,n, B},\nonumber\\
\beta\mu&=&-\left(\frac{\partial s}{\partial n}\right)_{\sigma^{\mu\nu},\epsilon, B},\qquad
\beta M=\left(\frac{\partial s}{\partial B}\right)_{\sigma^{\mu\nu},\epsilon, n}.\nonumber\\
\end{eqnarray}
To consider rotation (vorticity), the thermal vorticity tensor $\omega^{\mu\nu}\equiv \partial^{\mu}\beta^{\nu}-\partial^{\nu}\beta^{\mu}$, with $\beta^{\mu}\equiv u^{\mu}/T$ and its dual $\bar{\omega}^{\mu}=-\frac{1}{2}\epsilon^{\mu\nu\rho\sigma}u_{\nu}\omega_{\rho\sigma}$ are introduced. After normalization, the thermal vorticity satisfies $\omega_{\mu}\omega^{\mu}=-1$. Here, $\omega^{\mu}$ is the unit vector in the direction of rotation. The magnetic field is introduced by $B^{\mu}$ defined by $B^{\mu}=\frac{1}{2}\epsilon^{\mu\nu\rho\sigma}F_{\nu\rho}u_{\sigma}$. It is given in terms of a unit vector $b^{\mu}$, defined by $b^{\mu}\equiv \frac{B^{\mu}}{B}$ with $b_{\mu}b^{\mu}=-1$. The magnetic polarization vector is defined by $M^{\mu}$. In the rest frame of the fluid, where $B^{\mu}=(0,\bs{B})$ and $M^{\mu}=(0,\bs{M})$, we have $\bs{M}=\chi_{m}\bs{B}$, where $\chi_{m}$ is the magnetic susceptibility. To derive the energy-momentum tensor of ideal SVMHD, the expressions
$\mu^{\mu}=-\frac{1}{2}\epsilon^{\mu\nu\rho\sigma}u_{\nu}\mu_{\rho\sigma}$ and
$\sigma^{\mu}=-\frac{1}{2}\epsilon^{\mu\nu\rho\sigma}u_{\nu}\sigma_{\rho\sigma}$ with the following decomposition is used:
\begin{eqnarray}\label{appC6}
\mu^{\mu}&=&\mu_{\omega\|}  {\omega}^{\mu}+\mu_{b\|}b^{\mu}+\mu_{\perp}^{\mu},\nonumber\\
\sigma^{\mu}&=&\sigma_{\omega\|}  {\omega}^{\mu}+\sigma_{b\|}b^{\mu}+\sigma_{\perp}^{\mu}.
\end{eqnarray}
The subscripts $\|$ and $\perp$ denote the parallel and perpendicular directions with respect to the direction of the magnetic field and vorticity, $b^{\mu}$ and $\omega^{\mu}$. As it is shown in \cite{sedighi2024}, the energy-momentum tensor of an ideal, relativistic, and vortical fluid in the presence of external magnetic field is given by
\begin{eqnarray}\label{appC7}
T^{\mu\nu}&=&\epsilon u^{\mu}u^{\nu}-p_{\perp}\Xi^{\mu\nu}+p_{\times} \omega^{\mu}\omega^{\nu}+p_{\|} b^{\mu}b^{\nu}\nonumber\\
&&+\frac{B^{2}}{2}\left(u^{\mu}u^{\nu}-\Xi^{\mu\nu}-b^{\mu}b^{\nu}+\omega^{\mu}\omega^{\nu}\right)\nonumber\\
&&+2\mu_{b\|}\sigma_{\omega\|}b^{(\mu}\omega^{\nu)},
\end{eqnarray}
with three different types of pressure defined by
\begin{eqnarray}\label{appC8}
p_{\perp}&\equiv&p_{0}-\chi_{m}B^{2},\nonumber\\
p_{\times}&\equiv&p_{\perp}+\mu_{\omega\|}\sigma_{\omega\|},\nonumber\\
p_{\|}&\equiv&p_{0}+\mu_{b\|}\sigma_{b\|}.
\end{eqnarray}
Here, $\Xi^{\mu\nu}=g^{\mu\nu}-u^{\mu}u^{\nu}+b^{\mu}b^{\nu}+\omega^{\mu}\omega^{\nu}$.
In Sec. \eqref{sec3}, we compare \eqref{appC7} with the energy-momentum tensor \eqref{E7} which is derived by utilizing the Wigner function \eqref{N23}. In particular, we choose $\mu_{b\|}=\mu_{\omega\|}=0$.
%%%%%%%%%%%%%%%%%%%%%%%%%%%%%%%
\section{Proof of \eqref{E3}}\label{appD}
\setcounter{equation}{0}
%%%%%%%%%%%%%%%%%%%%%%%%%%%%%%%
In this appendix, we prove that the energy-momentum tensor of a rigidly rotating fermionic plasma is given by \eqref{N3}. We first outline the proof for flat spacetime, and then, we argue that in curved spacetime, by using the RNC approximation, the energy-momentum tensor is expressed by \eqref{N3}.
\par
We begin by considering the Lagrange density of a nonrotating QED plasma. It is given by \eqref{E1}, where the covariant derivatives
$\vec{D}^{\mu}$ and $\cev{D}^{\nu}$ act on $\psi$ and $\bar{\psi}$, respectively.
For a generic operator $\mathcal{O}$, we have $\bar{\psi}\cev{\mathcal{O}} = [\mathcal{O}\psi]^{\dagger}\gamma^{0}$. In flat spacetime, the covariant derivative is given by $D_{\mu}=\partial_{\mu}-i A_{\mu}$. The energy-momentum tensor associated with this Lagrangian is thus given by
\begin{eqnarray}\nonumber\label{appD1}
 		T^{\mu\nu} &=& \langle: \mathcal{T}^{\mu\nu} :\rangle \\\nonumber
		&=& \frac{i}{4} \langle : \bar{\psi} \left(\gamma^{\mu}\vec{D}^{\nu}+\gamma^{\nu}\vec{D}^{\mu}-\gamma^{\mu}\cev{D}^{\nu}-\gamma^{\nu}\cev{D}^{\mu}\right)\psi :\rangle \\
		&&  - \frac{i}{2}g^{\mu\nu} \langle : \bar{\psi} 	\left(\gamma^{\alpha}\vec{D}_{\alpha}-\gamma^{\alpha}\cev{D}_{\alpha}\right)\psi : \rangle.
	\end{eqnarray}
We insert the definition of the covariant derivative into \eqref{appD1} and consider, as an example, the first contribution $\langle : \bar{\psi} \gamma^{\mu}\vec{\partial}_{\nu}\psi : \rangle$. Inserting appropriate kets and bras, we first arrive at
    \begin{widetext}
    \begin{eqnarray}\label{appD2}
		\langle : \bar{\psi} \gamma^{\mu}\vec{\partial}_{\nu}\psi : \rangle = \int d^{4}y 	\frac{d^{4}p'}{(2\pi)^4}\frac{d^{4}p}{(2\pi)^4} \langle : \bar{\psi}|x+y/2\rangle \langle x+y/2|p'\rangle\langle p'| \gamma^{\mu}\vec{\partial}^{\nu}|p\rangle\langle p| x-y/2\rangle\langle x-y/2|\psi : \rangle.
\end{eqnarray}
Here, $\psi(x)=x|\psi\rangle$ is used. Introducing $P^{\nu} \equiv -i \partial^{\nu}$, we have $P^{\nu}|p\rangle=p^{\nu}|p\rangle$. Using the plane wave $\langle x|p \rangle= e^{ip\cdot x}$, we obtain
\begin{eqnarray}\label{appD3}
\langle : \bar{\psi} \gamma^{\mu}\vec{\partial}_{\nu}\psi : \rangle
=i \int \frac{d^{4}p}{(2\pi)^{4}}\gamma_{\beta\alpha}p^{\nu} \int d^{4}y e^{ip.y} \langle : \bar{\psi}_{\beta}(x+y/2) \psi_{\alpha}(x-y/2) : \rangle.
\end{eqnarray}
All the other terms in \eqref{appD1} can be determined similarly. After some computation, we arrive at
\begin{eqnarray}\label{appD4}
T^{\mu\nu} &=& -\left[ \frac{1}{2} \left(\delta_{\lambda}^{\mu}\delta_{\sigma}^{\nu}+\delta_{\lambda}^{\nu}\delta_{\sigma}^{\mu} \right) -g^{\mu\nu}g_{\lambda\sigma} \right] \int \frac{d^{4}p}{(2\pi)^4} \left(p^{\sigma}-A^{\sigma} \right) \gamma_{\beta\alpha}^{\lambda} \int d^{4}y e^{ip.y} \langle : \bar{\psi}_{\beta}(x+y/2) \psi_{\alpha}(x-y/2) : \rangle \nonumber\\
&=& \left[ \frac{1}{2} \left(\delta_{\lambda}^{\mu}\delta_{\sigma}^{\nu}+\delta_{\lambda}^{\nu}\delta_{\sigma}^{\mu} \right) -g^{\mu\nu}g_{\lambda\sigma} \right] \int \frac{d^{4}k}{(2\pi)^4} k^{\sigma} \gamma_{\beta\alpha}^{\lambda} \int d^{4}y e^{-ik\cdot y} \langle : \bar{\psi}_{\beta}(x+y/2) e^{iA\cdot y} \psi_{\alpha}(x-y/2) : \rangle \nonumber\\
&=& \left[ \frac{1}{2} \left(\delta_{\lambda}^{\mu}\delta_{\sigma}^{\nu}+\delta_{\lambda}^{\nu}\delta_{\sigma}^{\mu} \right) -g^{\mu\nu}g_{\lambda\sigma} \right] \int \frac{d^{4}k}{(2\pi)^4} k^{\sigma} \mbox{tr}(\gamma^{\lambda} W),
\end{eqnarray}
where a change of variable $p^{\sigma}-A^{\sigma} \equiv -k^{\sigma}$ is utilized to achieve the second equality.
\end{widetext}
In curved spacetime, the covariant derivative is defined by the horizontal lift, $D_{\mu}=\nabla_{\mu}-\Gamma_{\mu\nu}^{\lambda}y^{\nu}\partial_{\lambda}^{y}$. The momentum operator is then modified as $P_{\mu}\equiv -i\nabla_{\mu}$, with $\nabla_{\mu}$ defined in Sec. \ref{sec2}.
Since the additional terms in the horizontal lift does not affect the eigenstates of the momentum, we thus have
\begin{eqnarray}\label{appD5}
\langle p'| \vec{D}^{\nu}|p\rangle = i\langle p'| \vec{P}^{\nu}|p\rangle =i p^{\nu}(2\pi)^{4}\delta^{4}(p-p').\qquad
\end{eqnarray}
In addition, in curved spacetime, the expression $\langle x|p \rangle$ differs from its counterpart in flat spacetime. It is no longer a simple plane wave; instead, it depends functionally on the metric. However, within the framework of RNC, this expression can be locally approximated by a plane wave. Hence, by following the same steps as in the flat spacetime, we can approximate the energy-momentum tensor as it is given in \eqref{E3}. 
%%%%%%%%%%%%%%%%%%%%%%%%%%%%%%%
\section{Useful integrals}\label{appE}
\setcounter{equation}{0}
%%%%%%%%%%%%%%%%%%%%%%%%%%%%%%%
In Sec. \ref{sec3}, we used following integrals to arrive at \eqref{E6}
\begin{eqnarray}\label{appE1}
\int_{-\infty}^{+\infty} \frac{d^{2}k_{\perp}}{(2\pi)^{2}} e^{-\frac{\bs{k}_{\perp}^{2}}{eB}}&=& \frac{eB}{4\pi},\nonumber\\
\int_{-\infty}^{+\infty} \frac{dk_{i}}{(2\pi)} e^{-\frac{k_{i}^{2}}{eB}} k_{i} L_{n}^{1}\left( \frac{2\bs{k}_{\perp}^{2}}{eB}\right)&=& 0,\nonumber\\
\int_{-\infty}^{+\infty} \frac{d^{2}k_{\perp}}{(2\pi)^{2}} e^{-\frac{\bs{k}_{\perp}^{2}}{eB}} L_{n}\left( \frac{2\bs{k}_{\perp}^{2}}{eB}\right)&=& (-1)^{n}\frac{eB}{4\pi}, \nonumber\\
\int_{-\infty}^{+\infty} \frac{d^{2}k_{\perp}}{(2\pi)^{2}} e^{-\frac{\bs{k}_{\perp}^{2}}{eB}} \bs{k}_{\perp}^{2} L_{n}^{1}\left( \frac{2\bs{k}_{\perp}^{2}}{eB}\right)&=& (-1)^{n}\frac{(eB)^{2}(n+1)}{4\pi}.\nonumber\\
\end{eqnarray}
\end{appendix}

%%%% %%%%%%%%%%%%%%%%%%%%%

\end{document}